\documentclass[%
reprint,%
amssymb,amsmath,%
aip,apl,%
groupedaddress,%
]{revtex4-1}
\usepackage[dvipdfmx]{graphicx}
\usepackage{txfonts}
\usepackage{physics}
\usepackage{upgreek}    
\usepackage[colorlinks=true,linkcolor=blue]{hyperref}%
\newcommand{\micron}{$\upmu$m}
\newcommand{\degC}{$\mathrm{}^\mathrm{\circ}\mathrm{C}$}
\expandafter\ifx\csname package@font\endcsname\relax\else
 \expandafter\expandafter
 \expandafter\usepackage
 \expandafter\expandafter
 \expandafter{\csname package@font\endcsname}%
\fi
\begin{document}

\title{Characterization of X-ray charge neutralizer using carbon-nanotube field emitter}

\author{Shuhei Okawaki}
\author{Satoshi Abo}
\author{Fujio Wakaya}
\altaffiliation{Corresponding author: wakaya@stec.es.osaka-u.ac.jp}
\author{Hayato Yamashita}
\author{Masayuki Abe}
\author{Mikio Takai}
\affiliation{%
\vspace*{2mm}
Graduate School of Engineering Science, Osaka University, %
Toyonaka, Osaka 560-8531 Japan
}
\date{Dec. 9, 2015 submitted to Jpn. J. Appl. Phys; revised on March 31, 2016}

\begin{abstract}
An X-ray charge neutralizer using a screen-printed carbon-nanotube field emitter 
is demonstrated to show the possibility of a large-area flat-panel charge neutralizer, 
although the device dimensions in the present work are not very large.    
The X-ray yields and spectra are characterized to estimate the 
ion generation rate as one of the figures of merit of neutralizers.   
Charge neutralization characteristics are measured and show good 
performance. \\ \\
Journal reference: Jpn. J. Appl. Phys. \textbf{55}, 06GF10 (2016) \hspace{3mm}
DOI: 10.7567/JJAP.55.06GF10
\end{abstract}

\maketitle

\section{Introduction}

It is widely known that static electricity causes troubles 
not only in high-tech industries
but also in 
many fields including the manufacture of plastic, rubber, powder, and paper, 
and industries of textile, spinning, and printing,
meaning that the management of static electricity has been  
important in many industries for a long time.\cite{Noll1995,Murata2004}  
In the ULSI industry, 
the device dimensions become continuously small\cite{ITRS2013}, 
leading to a more fragile device against static electricity. 
In the industry of flat-panel displays, 
the screen size and pixel dimension become continuously large and small, 
respectively, 
meaning that the product becomes more sensitive to particle contaminations 
caused by static electricity.  
These outstanding trends in high-tech industries suggest 
that the management of static electricity becomes more critical presently.

Although material modification by antistatic additive doping or surface coating 
is effective for overcoming static electricity problems\cite{Noll1995},  
such a technique cannot always be adopted because, especially for electronic devices, 
an insulating material is necessary in many cases 
for realizing device functions.  
Charge neutralizers using air molecules ionized by corona discharge or 
soft X-ray irradiation are, therefore, often used for solving static electricity 
problems. \cite{Noll1995, Murata2004, Choi2004, Choi2005, Inaba1994, Inaba1996, Kawai2006}

Corona-discharge-type charge neutralizers have 
the following disadvantages, although 
they are widely used:\cite{Murata2004} 
(1) the charge balance of ionized air is not good, 
(2) the discharge process generates particles and causes contamination problems, 
(3) the density of ionized air is low on the target material surface, although it is high
      around the discharge electrodes, 
(4) the discharge process generates electromagnetic noise that may 
      cause troubles in the target devices to be neutralized. 
The X-ray charge neutralizer has advantages concerning all the above problems.  
Particularly when the X-ray source is shaped to a large-area flat panel, 
it should be useful for present large-area flat-panel devices 
because of the uniform ion density over a large area.  

A carbon nanotube (CNT) is one of the promising materials for 
electron field emitters owing to its high aspect ratio, high current tolerance, 
high mechanical strength, and high chemical stability.\cite{Iijima1991,DeHeer1995a,Collins1996}.  
Many applications using CNTs as field emitters, such as 
a field emission display,\cite{Wang2001,Choi2001}
a backlight unit for a liquid crystal display,\cite{Kim2005,Kim2009}
and an X-ray source,\cite{Sugie2001,Jeong2013,Manabe2013} are reported. 
A screen-printed CNT mat with appropriate post surface treatments 
is one of the best candidates for realizing 
a large-area field emitter\cite{Zhao2002, Zhao2003, Zhao2004, Sawada2003, Kanazawa2004, Was2005, Ohsumi2007, Oki2009, Takikawa2010, Manabe2012, Nitta2012}
and can be applied to a large-area X-ray 
source\cite{Manabe2013, Okawaki2015}.  
Although the X-ray generated at the large-area source may not be
very suitable for high-resolution X-ray imaging, 
it should be useful for realizing 
a large-area X-ray charge neutralizer for large-area 
flat-panel devices.  

In this study, an X-ray charge neutralizer using a screen-printed CNT cathode 
is demonstrated.  
Characterization results  show good performance for a charge neutralizer.

\section{Experimental methods}

\begin{figure}[t]
\centering
\includegraphics[width=80mm]{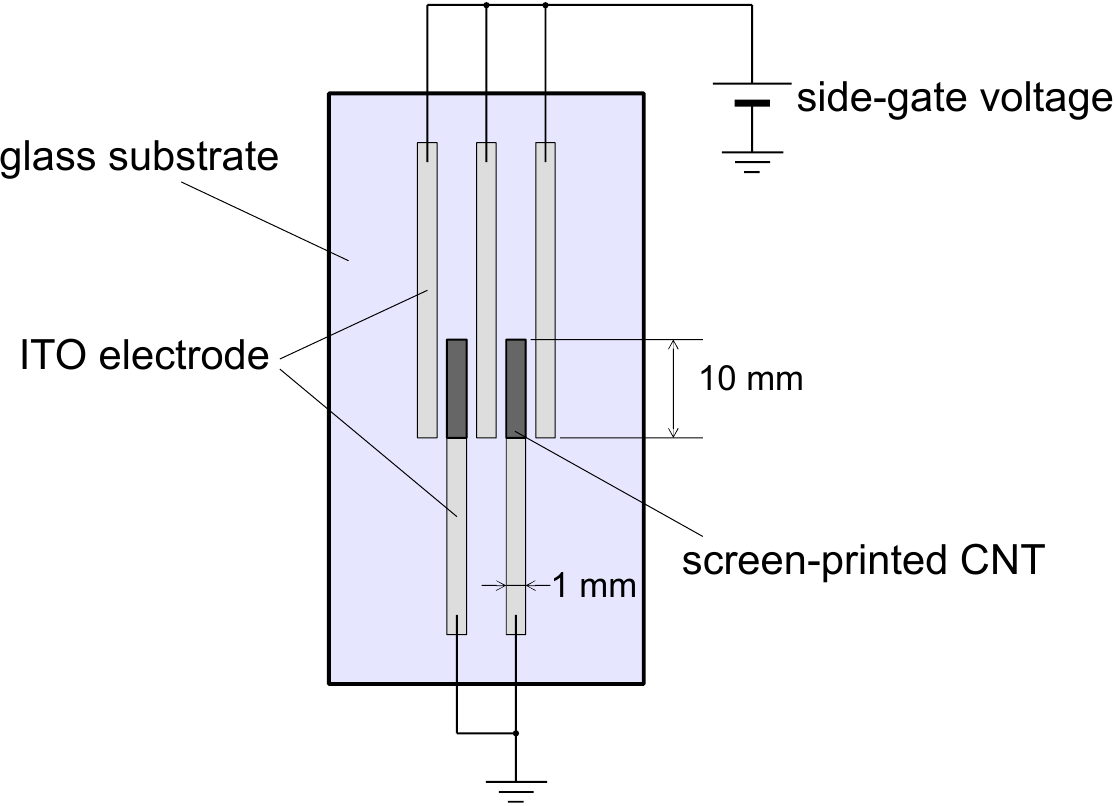}
\caption{(Color online) Schematic top view of screen-printed CNT cathode with side-gate electrodes.  
The bias setup for extracting electrons is also shown.}
\label{fig:fig1}
\end{figure}

To apply electron field emitters to an X-ray source,  a three-terminal configuration with 
emitters, gate electrodes, and an anode is necessary; such a configuration enables us to control separately the anode current from the anode voltage, leading to the separate control of 
the X-ray intensity and X-ray energy.  
An in-plane side-gate structure is adopted in the present work because 
it is easily applied to a large-area flat-panel emitter. 
The fabrication process for the emitter, which was already reported,\cite{Manabe2012, Nitta2012, Okawaki2015} 
is summarized as below. 
Indium-tin-oxide (ITO) electrodes of $30 \times 1$ mm$^2$, 
some of which were used as side-gate electrodes and 
others as back electrodes for CNT cathodes,   
were deposited on a glass substrate.  
A CNT paste was screen-printed on a part of the ITO electrodes with an area of 
$10 \times 1$ mm$^2$.  
The gap between the edges of the CNT cathode and side-gate electrode is 100 \micron.  
The schematic top view of the cathode is shown in Fig.~\ref{fig:fig1}.  
The effective area of the cathode, 
which is $\sim\!4 \times  10$ mm$^2$, 
is not very large
but it can easily be enlarged because  
screen printing can easily be applied 
to a large-area process.  
The tape-peeling surface treatment was performed to improve the electron emission property.\cite{Oki2009, Takikawa2010}

\begin{figure}[t]
\centering
\includegraphics[width=75mm]{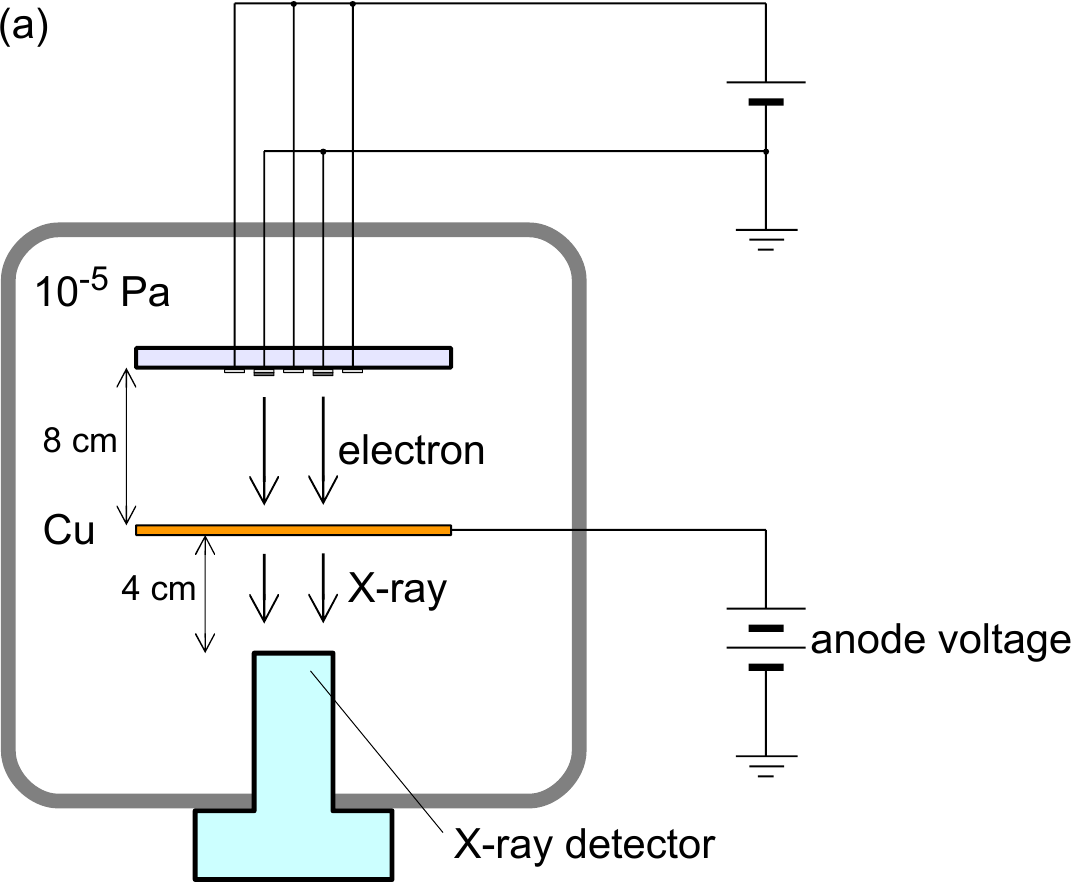}\\
\vspace{5mm}
\centering
\includegraphics[width=75mm]{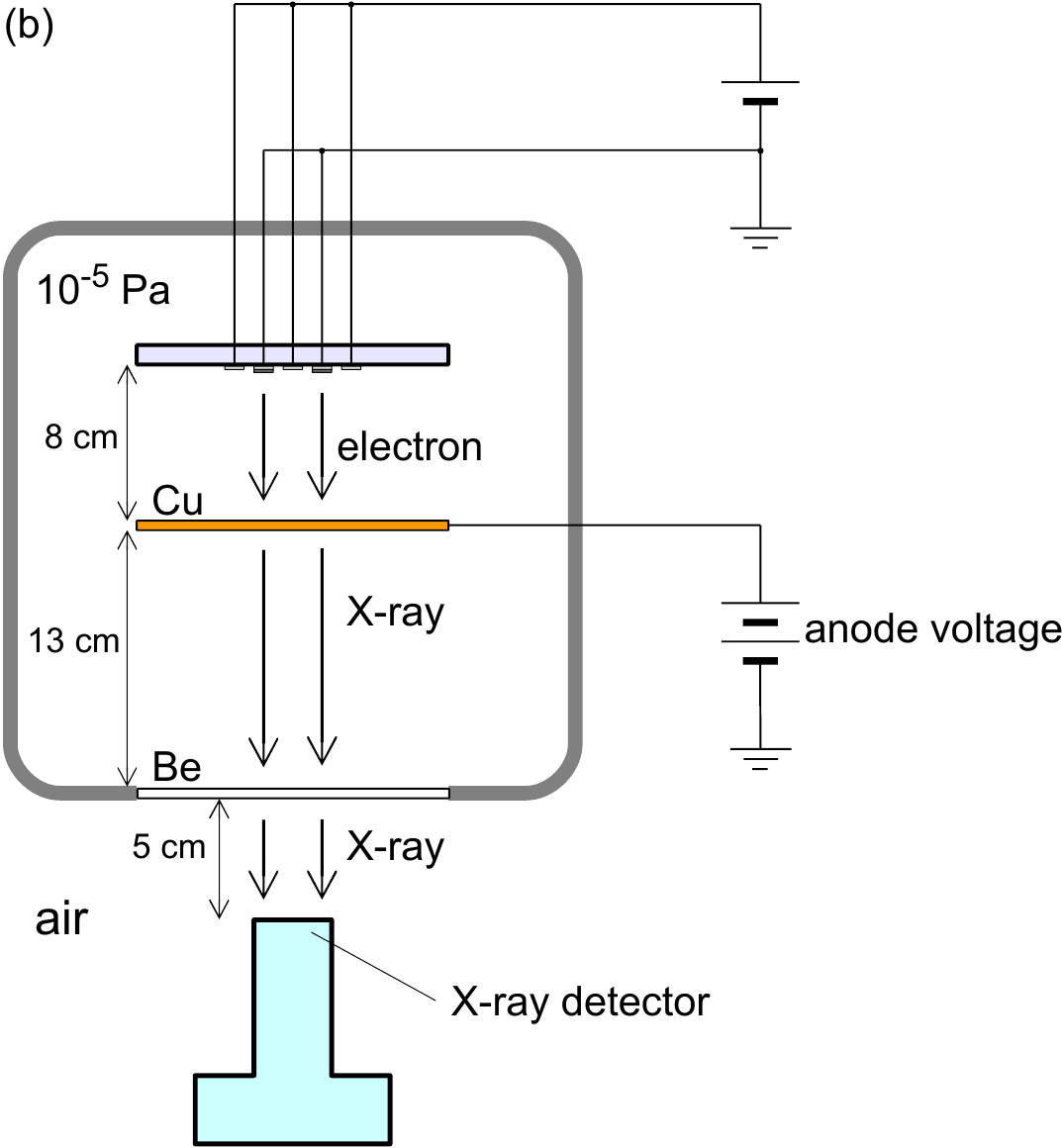}
\caption{(Color online) Experimental setup for generating and detecting X-rays.  
The X-ray detector is placed in vacuum (a) or in air (b). }
\label{fig:fig2}
\end{figure}

The experimental setup used to generate and detect X-rays is schematically shown 
in Fig.~\ref{fig:fig2}.  
In a vacuum of $\sim\!10^{-5}$ Pa, 
a 10-\micron-thick Cu thin film, placed 8 cm away from the emitter, 
was irradiated by field-emitted electrons
from the CNT cathode typically at 10 keV  to generate X-rays.  
To characterize the X-ray spectrum and yield in vacuum, 
the X-ray detector was set 4 cm away from the Cu film 
as shown in Fig.~\ref{fig:fig2}(a).  
To characterize X-rays in air, 
a 250-\micron-thick Be window was used and 
the X-ray detector was set in air 5 cm away from the Be window
shown in Fig.~\ref{fig:fig2}(b).  
The X-ray detection system used is Amptek XR-100CR/PX4 with a detector area of 
13 mm$^2$.

\begin{figure}[t]
\centering\includegraphics[width=65mm]{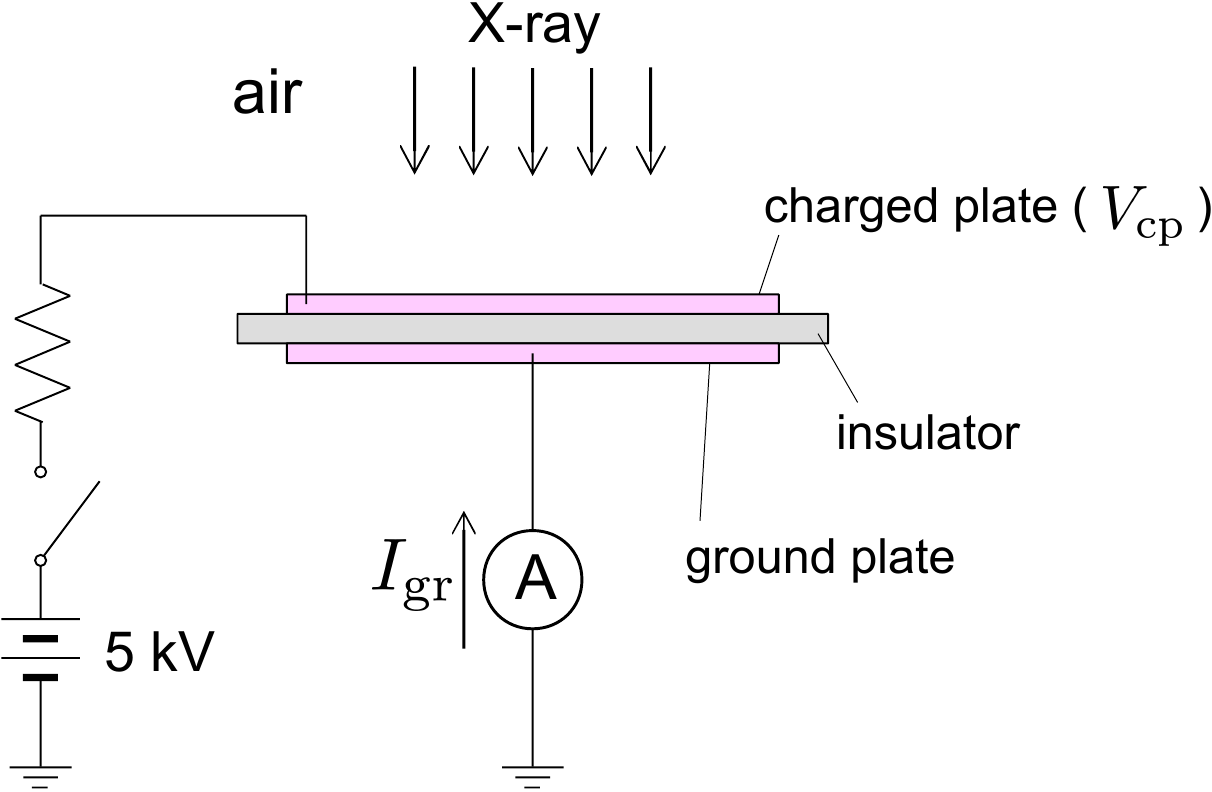}
\caption{(Color online) Schematic of charged plate monitor with measurement setup. }
\label{fig:fig3}
\end{figure}

Ion current was measured in air by a metal plate of 40 $\times$ 40  mm$^2$, 
placed 5 cm away from the Be window instead of the X-ray detector shown in 
Fig.~\ref{fig:fig2}(b). 
The bias voltage of the plate was kept constant at $-1$ kV during the measurement.   
This is not exactly the same as the real neutralization situation, because 
the voltage of the object material decreases during the real neutralization process.  
Such a technique, however, is often used for characterizing charge 
neutralizers.\cite{Murata2004} 
Charge neutralization performance as a function of time was 
characterized by using the charged plate monitor shown schematically 
in Fig.~\ref{fig:fig3} 
placed in air 5 cm away from the Be window.  
The area and capacitance of the charged plate monitor are 
25 $\times$ 25 mm$^2$ and  25 pF, respectively.   
The current $I_\mathrm{gr}$ defined in Fig.~\ref{fig:fig3} was measured as a function of time $t$,  since the high-voltage 
source was disconnected from the circuit.  
The electric potential of the charged plate, $V_\mathrm{cp} $, 
can be estimated as 
\begin{equation}
V_\mathrm{cp}(t) = V_\mathrm{cp}(0) - \frac{1}{C} \int_0^t I_\mathrm{gr}\, \dd{t} \, \textrm{,} \label{eq:eq1}
\end{equation}
with $V_\mathrm{cp}(0) = 5$ kV and $C = 25$ pF.  
The direct measurement of $V_\mathrm{cp}$ is difficult because 
the ion current and $I_\mathrm{gr}$ are typically $10^{-9}$ A at 
the electric potential of $10^3$ V as discussed in the following section, 
which means that 
a voltmeter input impedance of 
more than $10^{12}$ $\Omega$ 
is necessary.  This is the reason why the estimation using Eq.~(\ref{eq:eq1}) 
is adopted in this work.

\section{Results and discussion}

The anode current of the CNT cathode with side-gate electrodes was 
controlled by the side-gate voltage and reached 
the highest value of 300 $\upmu$A.    
The detailed field-emission properties were similar to those observed in the previous 
work\cite{Okawaki2015}.  

\begin{figure}[t]
\vspace*{5mm}
\centering\includegraphics[width=65mm]{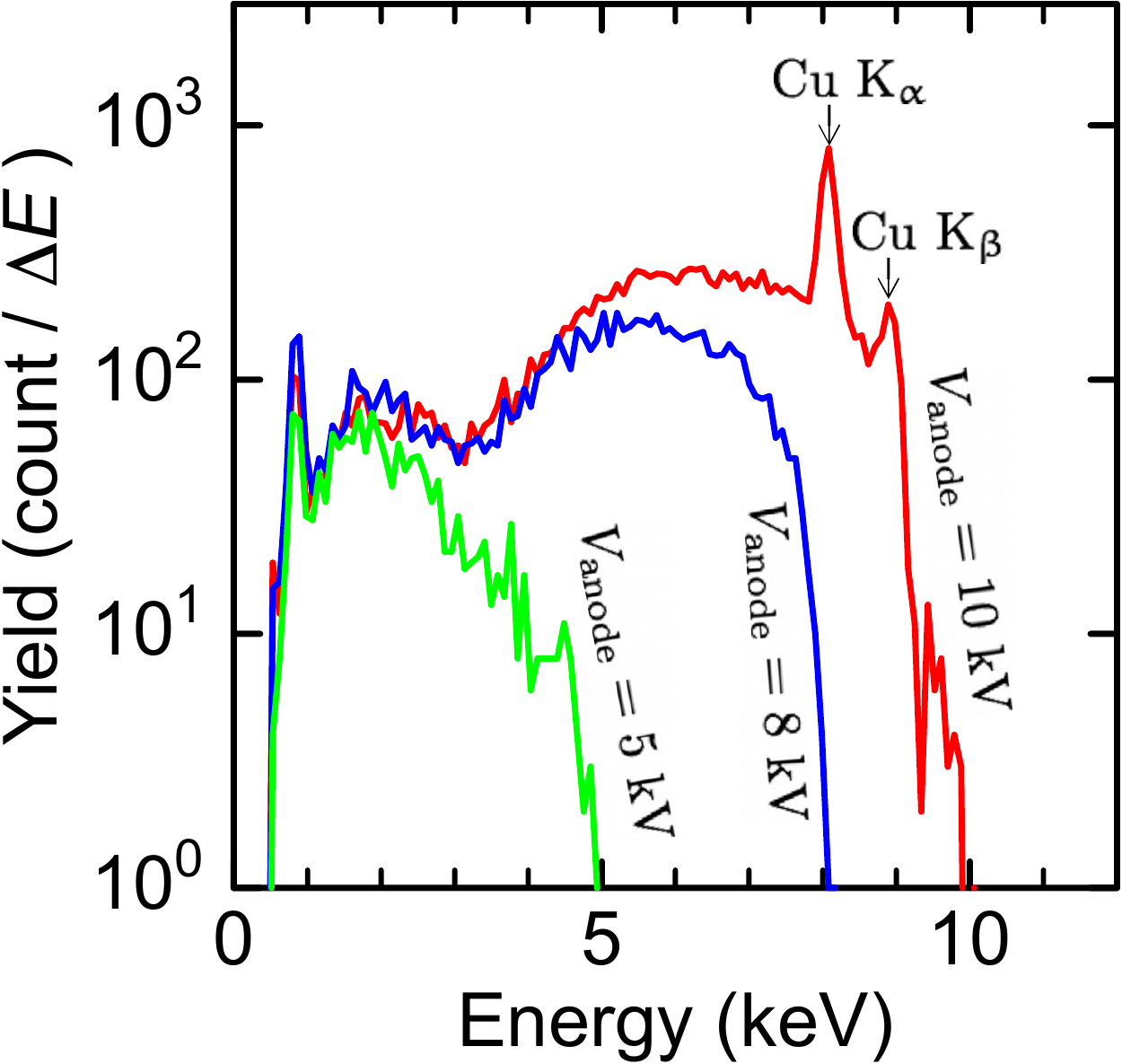}
\caption{(Color online) Energy spectra of X-rays measured in vacuum as in Fig.~\ref{fig:fig2}(a). 
The channel width $\Delta E$ of the multichannel analyzer is 89.9 eV.  
The anode current is fixed to 1 nA during the measurement.  
The measurement time for each spectrum is 60 s. 
}
\label{fig:fig4}
\end{figure}

Figure~\ref{fig:fig4} shows the X-ray spectra obtained in vacuum, 
the setup for which is shown in Fig.~\ref{fig:fig2}(a).  
The anode current during the measurement was reduced and kept at 1 nA, 
although it can be increased up to 
300 $\upmu$A as described previously, 
because the X-ray spectrum cannot be measured if it is very strong.  
The characteristic X-ray peaks of Cu were observed when 
the anode voltage was 10 kV.  For all anode voltages, 
the maximum energy of the bremsstrahlung X-ray  
corresponded to the anode voltage as expected.

\begin{figure}[t]
\centering\includegraphics[width=68mm]{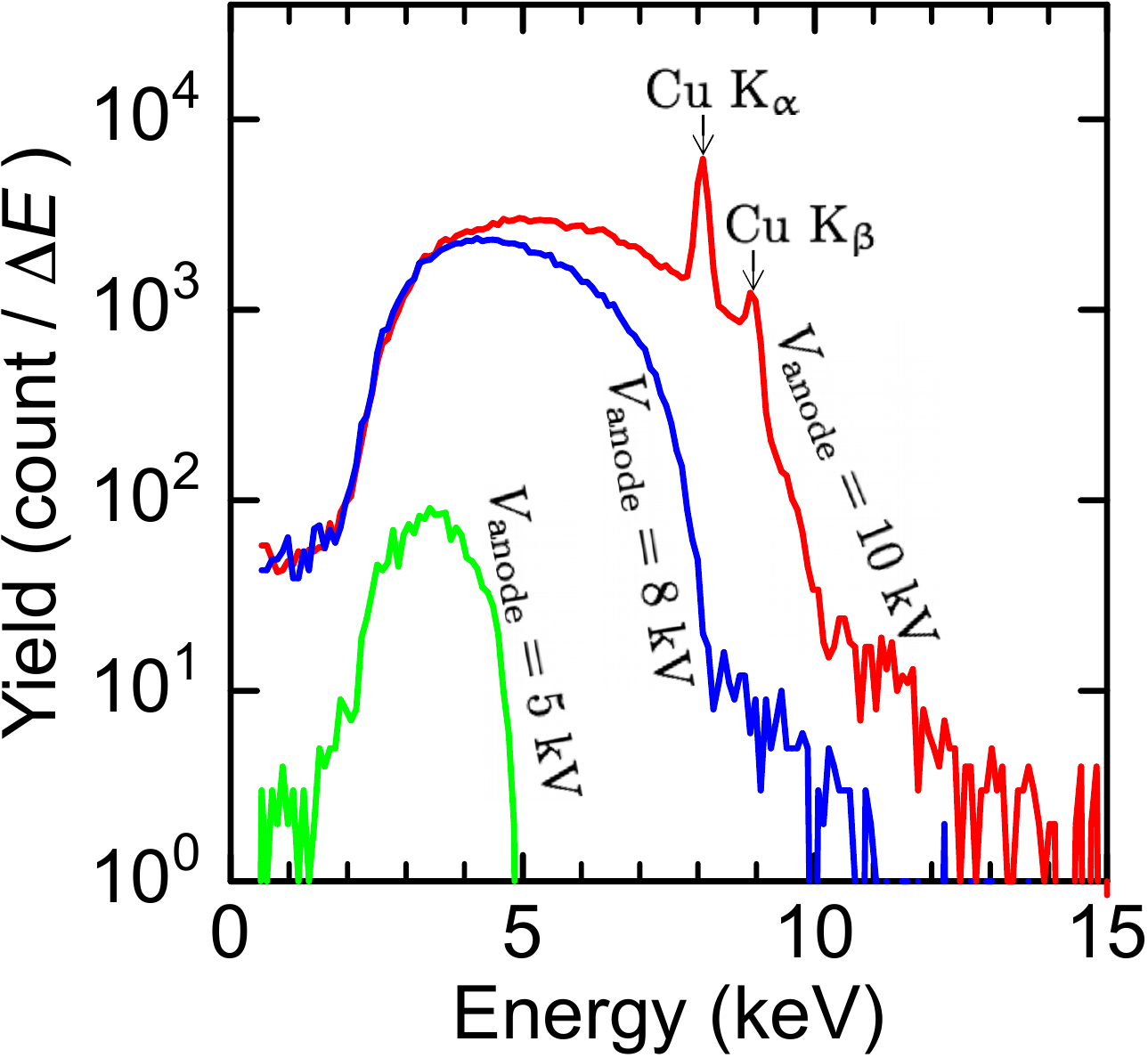}
\caption{(Color online) Energy spectra of X-ray measured in air as in Fig.~\ref{fig:fig2}(b). 
The channel width $\Delta E$ of the multichannel analyzer is 89.9 eV.  
The anode current is fixed to 400 nA during the measurement.  
The measurement time for each spectrum is 60 s. 
}
\label{fig:fig5}
\end{figure}

The X-ray spectra obtained in air are shown in Fig.~\ref{fig:fig5}.  
When the spectrum was measured in air,  
the X-ray was absorbed by air and its intensity decreased.  
The anode current was, therefore, increased to 400 nA, 
while the other parameters were maintained the same as in the in-vacuum measurement 
shown in Fig.~\ref{fig:fig4}.  
The maximum energies of the bremsstrahlung X-rays with 
anode voltages of 8 and 10 kV 
exceed the corresponding anode voltages not as expected.  
This is due to the double or multiple counting of photons 
during the time constant of the detection system.

\begin{figure}[t]
\vspace*{5mm}
\centering\includegraphics[width=50mm]{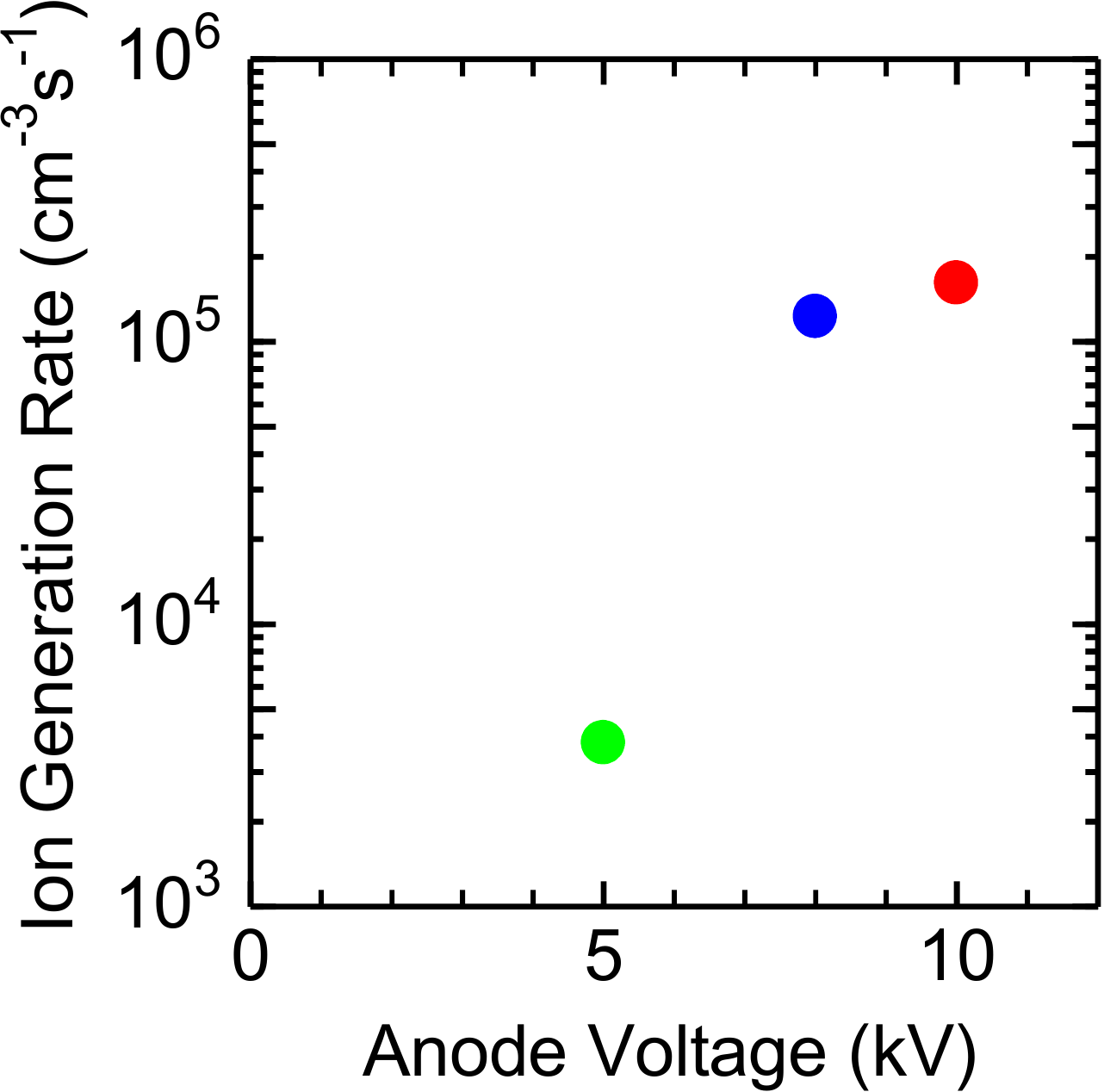}
\caption{(Color online) Estimated ion generation rate at 760 Torr and 20 \degC\ as a function of anode voltage. 
}
\label{fig:fig6}
\end{figure}

For the charge neutralizer, the ion generation rate $G$ is 
one of the important figures of merit, which can be estimated as
\begin{equation}
G(T,P) = \frac{1}{S\Delta t}
     \sum_i \frac{E_i}{W} Y(E_i) 
     \left[
     \frac{\mu_\mathrm{en}(E_i)}{\rho}
     \right] 
      \rho(T,P) \, \textrm{,}   \label{eq:eq2}
\end{equation} 
where 
$T$ and $P$ are the air temperature and pressure, respectively, 
$S$ is the area of the X-ray detector, 
$\Delta t$ is the measurement time for obtaining the spectrum, 
$E_i$ is the X-ray energy of the $i$th channel of the detector, 
$W$ is the average ionization energy of air, 
$Y(E_i)$ is the X-ray yield, 
$\mu_\mathrm{en}(E_i)$ is the X-ray absorption coefficient of air, 
$\rho$ is the mass density of air.  
Figure~\ref{fig:fig6} shows the ion generation rate estimated from 
the spectrum shown in Fig.~\ref{fig:fig5} using 
Eq.~(\ref{eq:eq2}) with 
$S=$ 13 mm$^2$, 
$\Delta t = 60$ s, 
$W=$ 34 eV, \cite{Iida2005,Macheret2002}
$\mu_\mathrm{en}/\rho$ from the database, \cite{Hubbell2004}
and $\rho = $ $1.205 \times 10^{-3}$ g/cm$^3$ at 760 Torr and 
20 \degC.\cite{Rikanempyou1990}
The maximum ion generation rate in Fig.~\ref{fig:fig6} is $1.6 \times 10^5$ 
cm$^{-3}$s$^{-1}$, corresponding to the ion concentration $C_\mathrm{ion}$ 
of $5.0 \times 10^5$ cm$^{-3}$.\cite{Inaba1996}
It is reported\cite{Inaba1996} that
$C_\mathrm{ion} = 10^{7}$ cm$^{-3}$ or 
$G=6.3 \times 10^{7}$ cm$^{-3}$s$^{-1}$ 
is enough for a charge neutralizer.  
The maximum $G$ estimated in the present work is 
two orders of magnitude lower than this value.  
The ion generation rate can, however, be increased by 
increasing the anode current, which was intendedly reduced 
from 300 $\upmu$A to 400 nA 
in order to obtain the X-ray spectrum shown in Fig.~\ref{fig:fig5}.  
Assuming that $G$ is proportional to the anode current, 
we can expect $G$ to be 750 times larger than those in the present work 
with an anode current of 300 $\upmu$A, 
the upper limit of the emitter used in the present work.  
This means that the charge neutralization performance of the 
present device is fairly good.

\begin{figure}[t]
\centering\includegraphics[width=55mm]{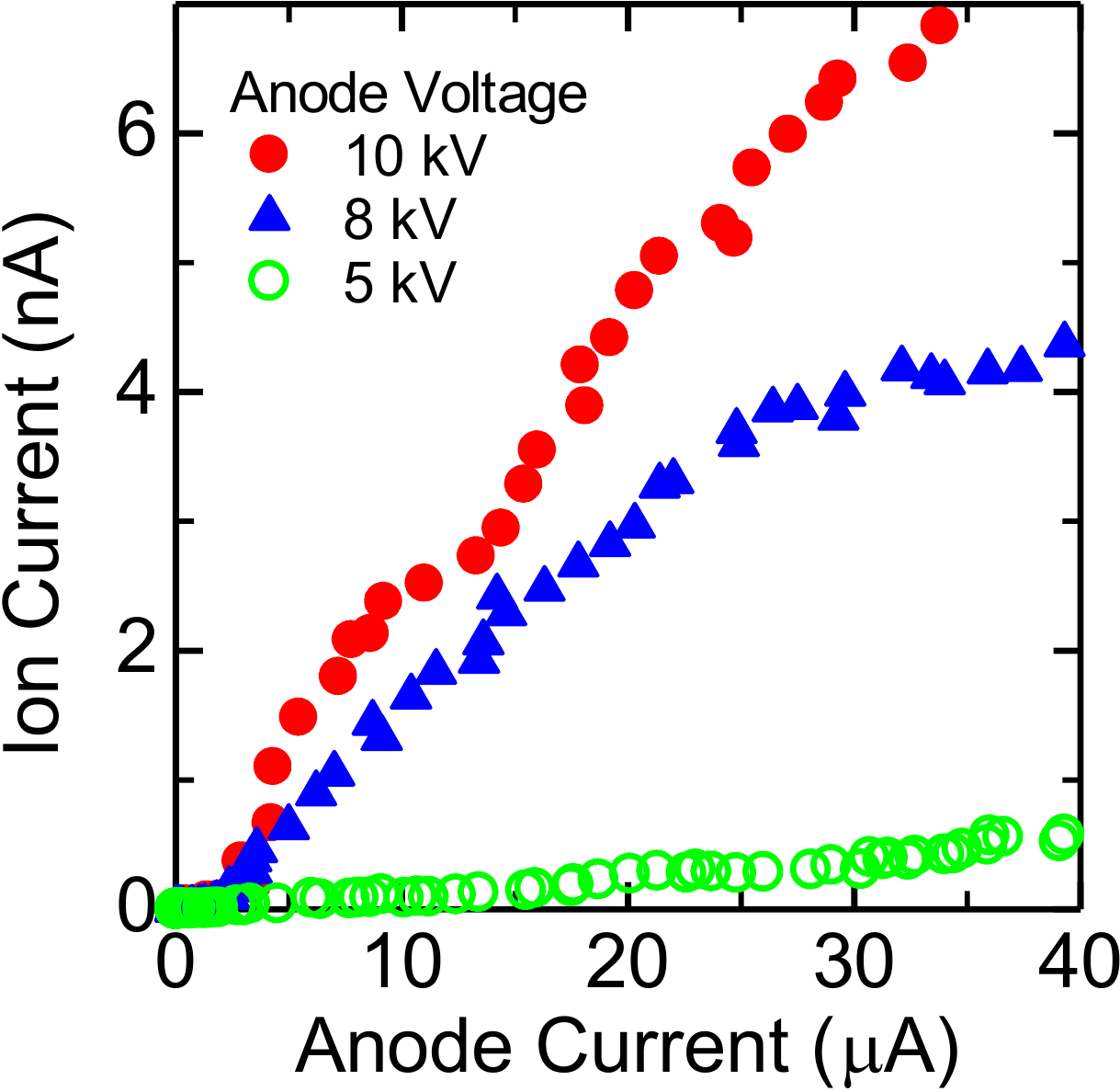}
\caption{(Color online) Ion current measured in the air as a function of anode current.  
}
\label{fig:fig7}
\end{figure}

The ion currents observed in air by using the metal plate biased 
at $-1$ kV 
are shown in Fig.~\ref{fig:fig7} as a function of the anode current.  
The ion current was roughly proportional to the anode current as expected.  
The absolute value of ion current with positive bias should be similar because 
it is considered that the ion balance of the X-ray charge neutralizer is 
better than that of the corona-discharge-type 
neutralizer\cite{Murata2004}, although the measurement with 
positive bias was not performed owing to the voltage source restriction.  

\begin{figure}[t]
\vspace*{5mm}
\centering\includegraphics[width=55mm]{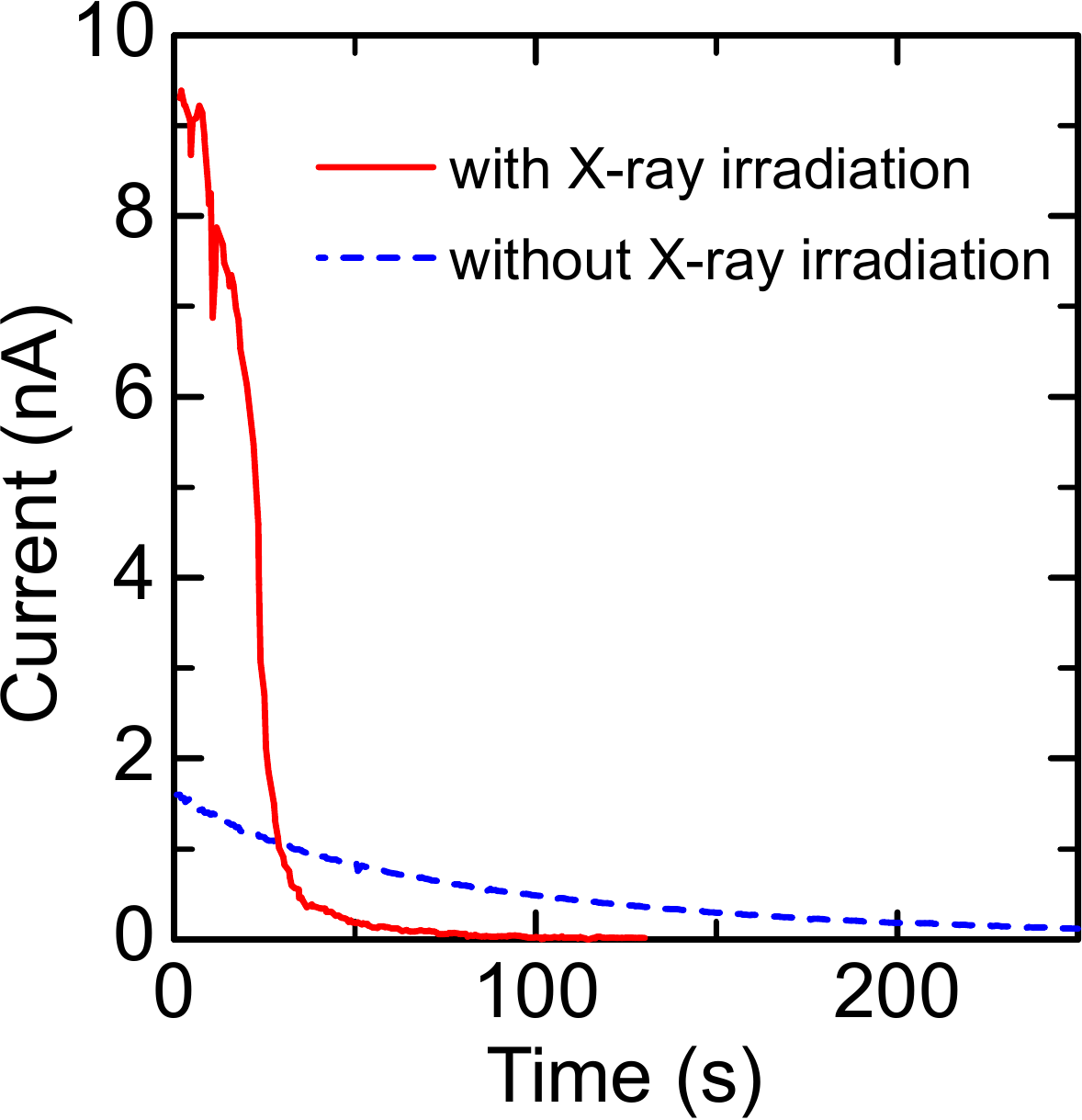}
\caption{(Color online) Current measured by the charged plate monitor shown in Fig.~\ref{fig:fig3}. 
The distance between the Be window and the charged plate is 5 cm.  
The anode voltage and anode current were 10 kV and 300 $\upmu$A, respectively, 
during the measurement.  
}
\label{fig:fig8}
\end{figure}

\begin{figure}[t]
\centering\includegraphics[width=55mm]{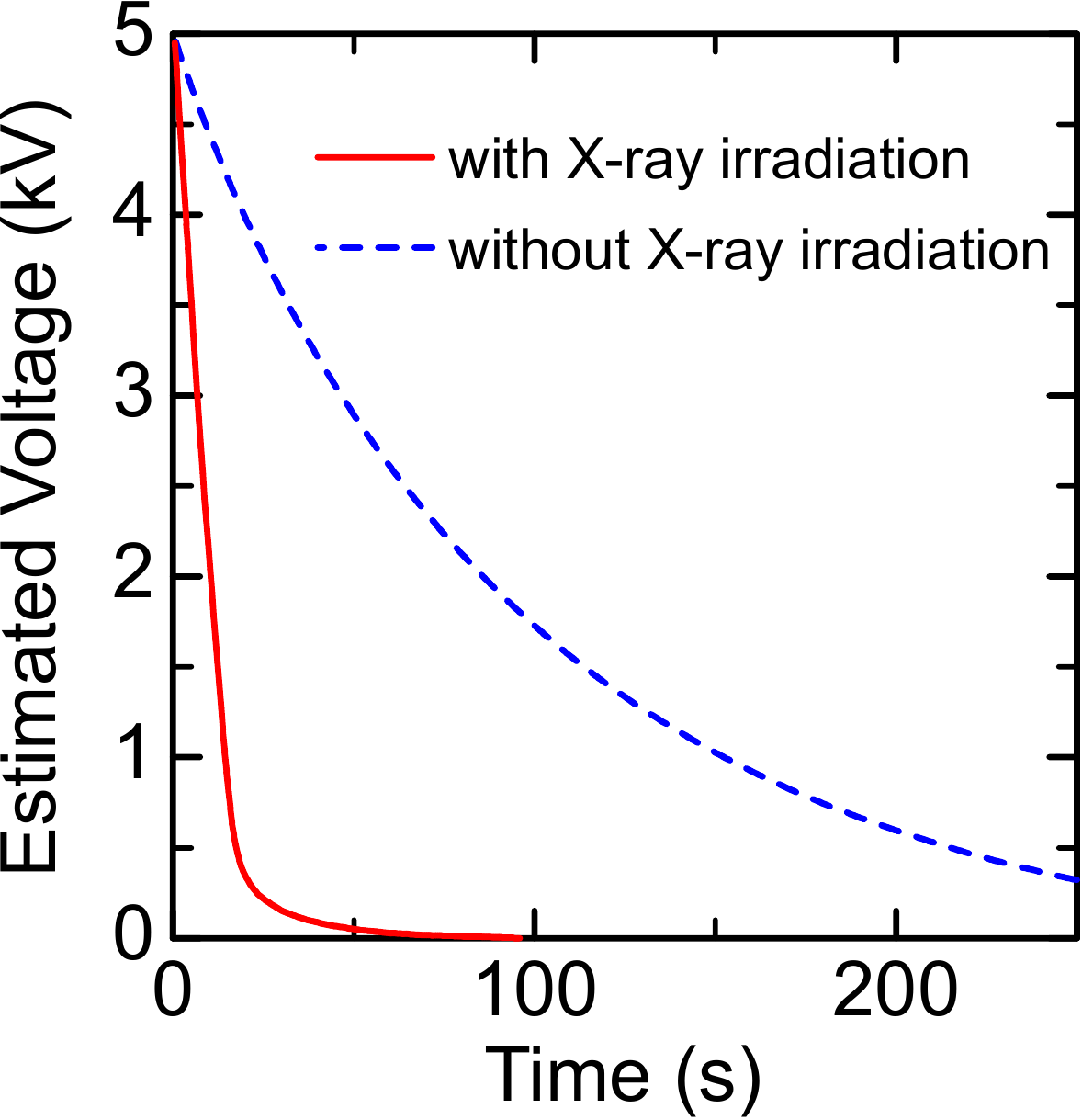}
\caption{(Color online) Voltage at the charged plate as a function of time 
estimated from the observed current shown in Fig.~\ref{fig:fig8}. 
}
\label{fig:fig9}
\end{figure}

The measured current to the ground plate $I_\mathrm{gr}$, defined in 
Fig.~\ref{fig:fig3}, is shown in Fig.~\ref{fig:fig8}.   
For this measurement, the anode current was increased and fixed to 
300 $\upmu$A to obtain the highest neutralization performance.  
When the current $I_\mathrm{gr}$ shown in this figure is 
used with Eq.~(\ref{eq:eq1}), $V_\mathrm{cp} \ne 0$ after a sufficiently long time.  
This is probably due to the unreliable value of the capacitance $C$.   
If $C$ is assumed to be  $\sim\! 10$\% larger than 25 pF, 
$V_\mathrm{cp} = 0$ after a sufficiently long time.  
The following normalization was, therefore, used 
instead of Eq.~(\ref{eq:eq1}) to avoid the problem: 
\begin{equation}
V_\mathrm{cp}(t) = 
V_\mathrm{cp}(0) 
\left(
 1-
\frac{\int_0^t I_\mathrm{gr}\, \dd{t}} 
{\int_0^\infty I_\mathrm{gr}\, \dd{t} }
\right)
\, \textrm{.} \label{eq:eq3}
\end{equation}
The resulting estimated electronic potentials at the charged plate 
with and without X-ray irradiation are shown in Fig.~\ref{fig:fig9}. 
The decay observed without X-ray irradiation should be due to the 
natural neutralization of the charged plate interacting with the 
surrounding air.  
If the decay time is defined as the time at which the 
voltage becomes $1/10$ of the initial value, 
it is estimated to be 17.5 and 215 s with and without X-ray irradiation, respectively.  
The decay time of 17.5 s with X-ray irradiation is not very good 
compared with the reference\cite{Inaba1996}, 
but can be improved by increasing the X-ray intensity.    
The X-ray intensity can be increased 
by decreasing 
the distances between the Cu target and the Be window and 
between the Be window and the object material to be neutralized, 
which are 13 and 5 cm, respectively, in the present work for 
the preliminary demonstration.

\section{Conclusions}

An X-ray charge neutralizer was demonstrated by using 
a screen-printed CNT field emitter.  
The effective area of the emitter is $\sim\! 4 \times 10$ mm$^2$.  
This is not very large and is almost a point source because the 
distances in the present work between the emitter and the Cu target,  
between the Cu target and the Be window, and 
between the Be window and the object material to be neutralized 
are 8, 13, and 5 cm, respectively.  
To realize a large-area flat-panel source is not difficult because 
screen printing can be easily applied to a large-area process.  
Charge neutralization characteristics were measured   
and showed good performance even under such an almost point-source condition, 
suggesting that the performance is much improved when 
a vacuum-sealed large-area flat-panel charge neutralizer is realized 
using the screen-printed CNT field emitter.  

\begin{acknowledgments}
This work was partially supported by JSPS KAKENHI 
Grant Number 23360022.  
\end{acknowledgments}

\bibliographystyle{aipnum4-1}

\end{document}